\documentclass[12pt]{article}
\usepackage{latexsym,epsfig,color}
\usepackage{colordvi}
\usepackage{amsmath}
\newcommand{\be}{\begin{equation}}
\newcommand{\ee}{\end{equation}}

\def\Grn{\Cyan}
\def\Grn{\Black}
\def\Prp{\Magenta}
\def\Prp{\Black}
\def\pd{\partial}

\def\pd{\partial}

\def\taup{\tau_{\parallel}}
\def\taupzero{\tau_{\parallel}}
\def\kapl{\chi_{\parallel}}
\def\kapp{\chi_{\perp}}

\def\hline{\centerline{\Red{\rule{6in}{1.0pt}}}}
\def\Alf{Alfv\'en }
\def\rfig#1{Fig.\ref{fig:#1}}

\def\req#1{(\ref{eq:#1})}
\def\res#1{Section \ref{sec:#1}}

\def\macname{hankrs2}
\def\macname{hank_strauss}
\def\fignf12{./pix}

\def\figdircc1{/Users/{\macname}/Documents/progs/m3dc1/plots}

\begin{document}
\begin{center}
{\it The following article has been submitted to Physics of Plasmas.} \\
{\Large{\bf Thermal Quench in ITER {Locked Mode}  Disruptions }} \\ 
H. Strauss  \\ 
 HRS Fusion, West Orange NJ, USA 07052 \\
email: hankrs2@gmail.com
\end{center}

\abstract{
Simulations and theory are presented of an ITER locked mode thermal quench (TQ).
In present experiments,
locked mode disruptions have a long precursor phase, followed by a
rapid termination and thermal quench, which can be identified with a resistive
wall tearing mode (RWTM).  In ITER, the RWTM will be slowed by the highly conductive
vacuum vessel. The rapid termination might be absent,
and the plasma could remain in the precursor phase.
If the edge temperature is in the collisional regime, 
the TQ would proceed on a long timescale,
limited by the RWTM to almost $100ms.$
This is  an important self mitigating effect.
}
\section{Introduction} \label{sec:intro}

Simulations and theory are presented of an ITER locked mode thermal quench (TQ).
Locked mode disruptions are the most common type in JET \cite{devries}.
In present tokamaks, locked mode disruptions have a long
precursor phase with moderate thermal loss
caused by tearing modes.
 This is  followed by a rapid TQ termination, which
is seen in JET \cite{devries} , DIII-D \cite{sweeney} and other devices.
The fast termination phase has rapid growth of magnetic perturbations
and abrupt  loss of thermal energy. A recent study \cite{rwtm21} has 
identified the TQ termination with a resistive wall tearing mode (RWTM).
In ITER \cite{iter}, the RWTM will be much slower
than in JET and other present devices, and the termination phase might be absent. The
thermal quench time might be much longer, if the edge temperature is in the collisional 
regime.   The need  for disruption
mitigation \cite{lehnen} by radiation
\cite{izzo,ferraro,nardon}
and runaway electron prevention \cite{reux}  might  be be substantially reduced. 

The growth rate of the RWTM is
\cite{rwtm21,finn95} 
\be \gamma \tau_A =  c_0 S^{-1/3} S_{wall}^{-4/9},\label{eq:gamma} \ee
where $S$ is the Lundquist number,
$S_{wall} = \tau_{wall} / \tau_A, $ 
where $\tau_{wall}$ is the resistive wall  magnetic penetration time,
$\tau_A = R / v_A$ is the \Alf time, and $R$ is the major radius.
The constant $c_0$ is given by simulations and theory \req{c0}.

It was shown \cite{rwtm21} that the TQ time $\tau_{TQ}$ is given by the
smaller of $1/\gamma$ or the parallel thermal transport time 
\be {\tau}_{TQ}  \approx \left( \frac{1}{ \gamma} , \taupzero \right)_{min}  \label{eq:adhoc} \ee
where 
\be \taupzero = \frac{a^2}{\kapl b_{n 0}^2}, \label{eq:taupar} \ee
$\kapl$ is the parallel thermal diffusivity in the plasma edge region, 
$b_n$ is the root mean square amplitude of magnetic
perturbations normal to  the plasma boundary,
\Grn{$b_{n 0}$ is the precursor amplitude of $b_n$
when the RWTM is negligible,}
and  $a$ is the minor radius  in the midplane.

\res{simulations} describes simulations in ITER geometry, initialized 
with an equilibrium whose evolution resembles a locked mode state. 
It will be shown that the RWTM can exist in ITER, but it is slowly growing
for the ITER value of $S_{wall}.$ 
Simulations
are done with a range of $S_{wall}$ values,
to verify 
the scaling of the RWTM growth rate and $\tau_{TQ}$ with $S_{wall}$,  and to obtain
the value of $c_0$ in \req{gamma} and $b_n$ in \req{adhoc}.

In \res{theory}, an analytic expression for the RWTM growth rate  is
compared with simulations. The analytic model of the TQ time \req{adhoc} is obtained
and also compared with the simulations.

In \res{params}, $\tau_{TQ}$  is calculated with realistic parameters using  \req{adhoc}. 
A model thermal conductivity is introduced  with collisionless and collisional
limits. The TQ time is
found as a function of edge temperature,
using values of $c_0$ and $b_n$ from the simulations, as well as $b_n$ values 
predicted from experimental data.

The value of $\tau_{TQ}$ can vary widely in ITER, depending on the
edge temperature and magnetic perturbation amplitude. 
If the amplitude of the edge magnetic perturbations is taken as the value found in
the simulations, $\tau_{TQ} > 10ms$ for any reasonable edge temperature $T.$
If the magnetic perturbation amplitude is $2$ times the simulation value, 
as suggested by a model based on experimental data \cite{devries},   the
edge temperature must be $T < 450eV$ to have $\tau_{TQ} > 10ms$.  
If the edge temperature is collisional
the TQ time can be tens of
$ms$ for any reasonable magnetic perturbation amplitude, up to almost $100ms,$
limited by a RWTM.

Discussion and conclusions are presented in \res{conclusion}.

\section{\bf ITER Thermal Quench Simulations } \label{sec:simulations}

\begin{figure}
 \begin{center}
   \includegraphics[height=5cm]{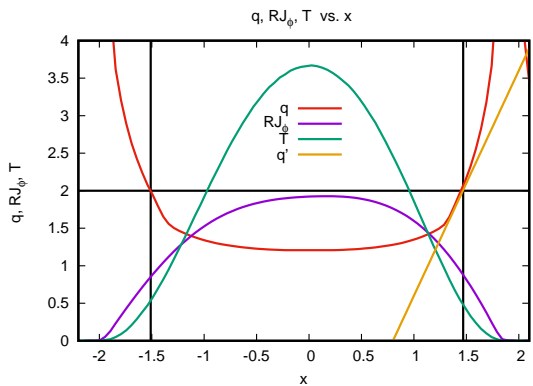} 
\end{center}
\caption{\it
  Initial profiles of $q$, toroidal current density $RJ_\phi$, and temperature  $T$
 as functions of $R - R_0$ in $m$, with $ Z = 0,$ and magnetic axis $R_0.$ 
 The straight line measures $q^\prime.$
}
\label{fig:initial}
\end{figure}

Simulations with M3D \cite{m3d} were performed to  examine the 
dependence of the TQ time on $S_{wall}.$
The simulations 
 are initialized with an   equilibrium with
 inductive Scenario 2 15 MA initial state \cite{iter},
formerly known as   ITER FEAT 15MA with
$li = 1.27,$ $\beta_{p} = 0.0079,$ and $q_0 = 1.2.$
The same equilibrium reconstruction was used in \cite{iter2018,fec20}.
Initial profiles of $q$, toroidal current density $RJ_\phi$, and temperature  $T$ 
are shown in \rfig{initial}, as functions of $R - R_0$ in $m$ through the magnetic axis
at $R_0,$ with $Z = 0.$ A straight line is fit to calculate  $q^\prime$ at the $q = 2$
surface. The density is assumed constant, so pressure $p \propto T.$ 
The profiles resemble locked mode profiles, with small $J\phi$ \cite{schuller} 
and $T$ outside the $q = 2$ surface. 
Typically locked mode disruptions have edge
cooling \cite{pucella} as a precursor.

The ITER vacuum wall is assumed to be a resistive  wall. The first wall is
assumed to have much higher resistivity. The walls are indicated in \rfig{iterw3d}, \rfig{iterw3d2}.


The simulations have initial Lundquist number $S = 10^6$ on axis, and 
 $100 \le S_{wall} \le 10^5.$ 
The  parallel thermal conductivity
is  $\chi_\parallel  = 10 R^2 / \tau_A $, 
and the perpendicular thermal conductivity is
$ \chi_\perp   = 10^{-4} a^2 / \tau_A. $
The choice of $\chi_\perp$ is
unrealistically  large, but it is constrained by the need to maintain numerical
stability. It is overwhelmed by parallel thermal conduction.  
The parameter values are not critical. They
serve to verify the scalings \req{gamma},\req{adhoc}, which can then be applied
in \res{params}  with realistic parameters.

\begin{figure}
 \begin{center}
   \includegraphics[height=5cm]{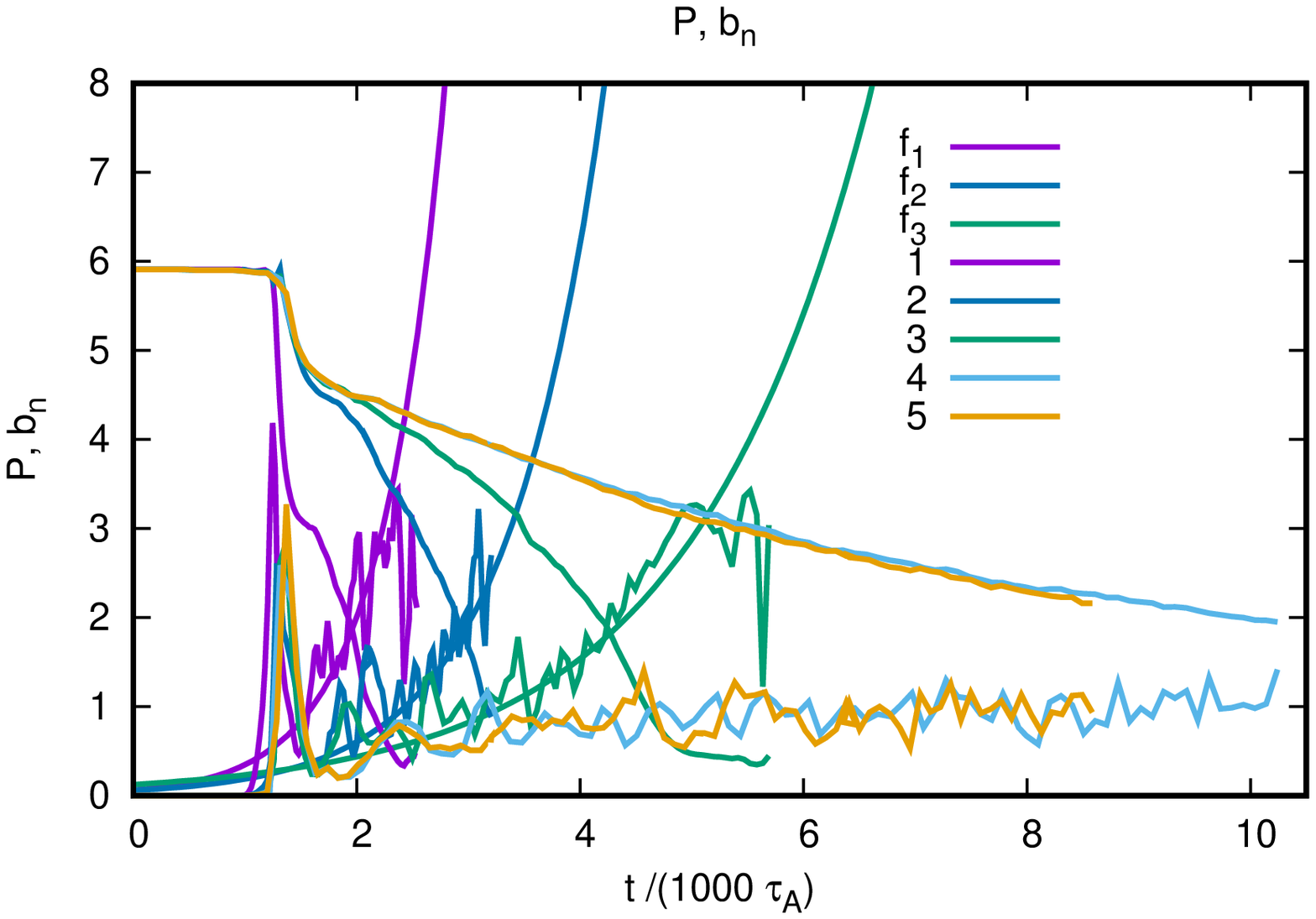}(a)
   \includegraphics[height=5cm]{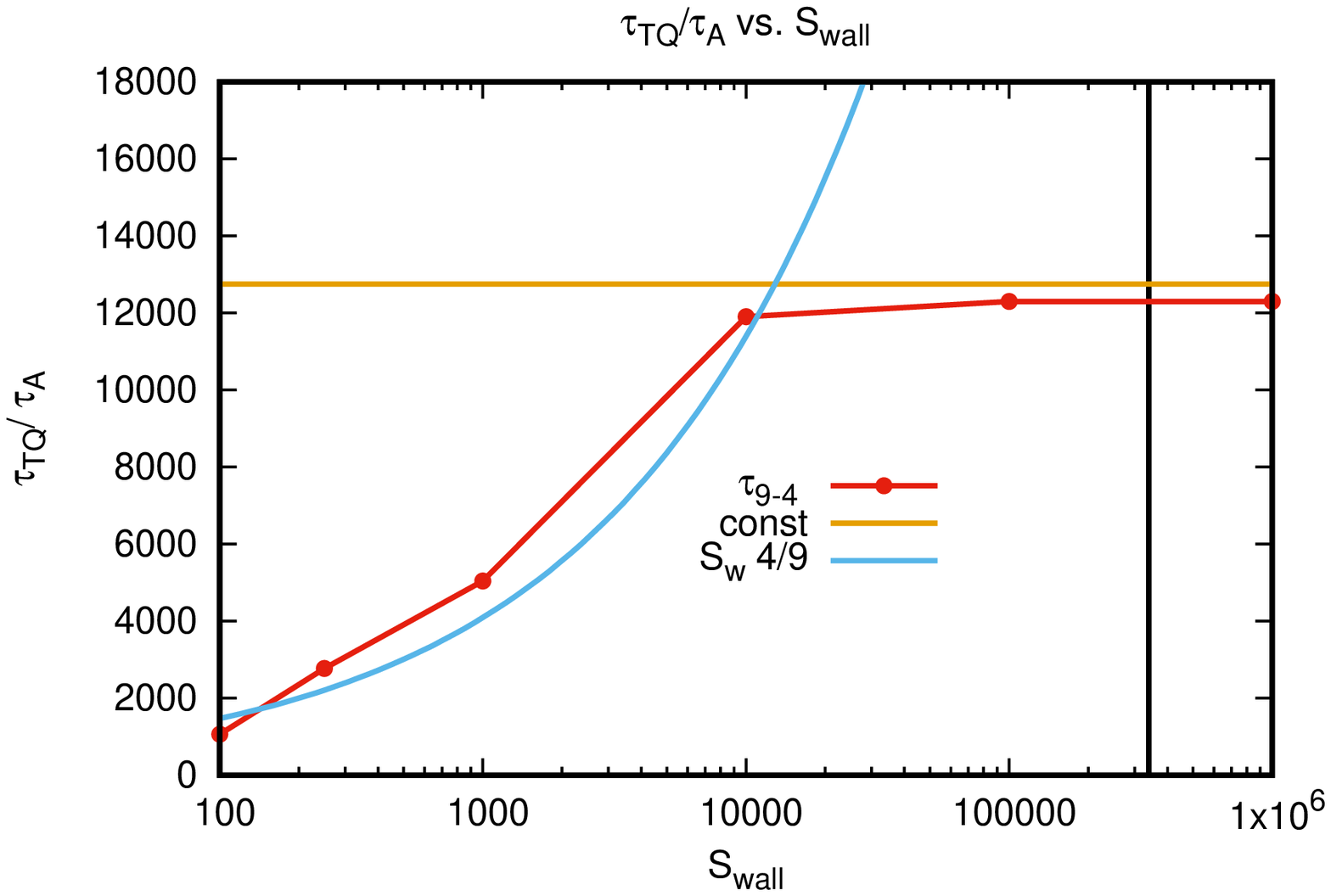}(b)
\end{center}
\caption{\it
(a)  $P$ and $b_n$ as a function of  $t/(1000 \tau_A),$ 
along with exponential fitting functions $f_1$ with $S_{wall} = 100$,
                             $f_2$ with $S_{wall} = 250$, and
                             $f_3$ with $S_{wall} = 10^3$.
The $P$ and $b_n$ curves have the same numbering as the fitting functions,
along with $4$ with $S_{wall} = 10^3$ and $5$ with $S_{wall} = 10^5$.
As $b_n$ increases in time, $P$ falls more rapidly.
(b) $\tau_{TQ}/\tau_A$ vs.  $S_{wall}.$
The fits are  $\propto S_{wall}^{4/9}$ and constant.
The vertical line is the ITER $S_{wall}.$ 
}
\label{fig:iterwall}
\end{figure}

\rfig{iterwall}(a) shows  
simulations  done for several values of $S_{wall}.$ 
The curves are  labelled $1$ for $S_{wall} = 100,$
                         $2$ for $S_{wall} = 250,$
                         $3$ for $S_{wall} = 10^3,$
                         $4$ for $S_{wall} = 10^4,$
                    and  $5$ for $S_{wall} = 10^5.$
The volume integral of the  pressure $P$ is shown in arbitrary units, 
as a function of time in $1000 \tau_A$ units.
The pressure decreases more slowly  as $S_{wall}$ increases. 
For $S_{wall} \ge 10^4,$ the decrease of total pressure
is independent of $S_{wall}.$ 
The decay of the pressure in \rfig{iterwall}(a) appears to involve three timescales. 
First is a fast decay of the pressure profile, which decreases by about 15\%, 
\Grn{for $S_{wall} \ge 250.$ }
This is due to a large internal kink, which 
produces a turbulent state. The turbulence  decays rapidly to a lower amplitude,
and along with a $(2,1)$ tearing mode,  causes relatively slow decay of $P$. 
There is a third faster phase, associated with the growth of a RWTM. 

The perturbed normal magnetic field at the wall, $b_n,$ 
is also shown. 
Here $b_n$ is defined as the
surface average along the first wall of the root mean square 
of the normal component of the
perturbed, asymmetric magnetic field  $\tilde{B}_n$, divided by the toroidal field $B_T$
on axis,
$b_n =  (2\pi L)^{-1/2}[\oint d\phi \oint d l (\tilde{B}_n / B_T)^2 ]^{1/2}$ where
$L = \oint d l.$
The units of $b_n$ are $10^{-3}$. The simulations with $S_{wall} \le 10^3$ have
maximum $b_n \approx 3\times 10^{-3},$ while the simulations with $S_{wall} \ge 10^4$ have
maximum $b_n \approx 1\times 10^{-3}.$ 

Also shown are
 exponential 
 fitting functions $f \propto \exp(c_0 S^{-1/3} S_{wall}^{-4/9}  t / \tau_A),$  
with the same $c_0 S^{-1/3}$ and
subscripts corresponding to the $S_{wall}$  numbering.
The fit yields the growth rate \req{gamma} 
 with $c_0 = 0.51.$ 
Here the value of $S$ was estimated from 
\rfig{initial}, which  shows that $T_s / T_{max}  = 0.5 / 3.5$, 
{where $T_s$ is the value of $T$ at the $q = 2$ surface.}
Then $S = (T_s / T_{max})^{3/2} S_{max} $, where $S_{max} = 10^6.$


\rfig{iterwall}(b) collects the $\tau_{TQ}$ data as a function of
$S_{wall}.$ The TQ time is measured as the
time difference $(t_{40} - t_{90})/.5,$ where $t_{90}$ is the time at which the
temperature is $90\%$ of its maximum value, and $t_{40}$ is the time when it has
$40\%$ of  its maximum value. 

A  fit $\tau_{TQ} \propto \exp(S_{wall}^{4/9})$  in \rfig{iterwall}(b) 
 yields $\gamma\tau_{TQ} = 1.16 \approx 1,$ with $\gamma$ from \rfig{iterwall}(a). 
The data in \rfig{iterwall}(b)  is approximately fit by formula \req{adhoc}. 
The vertical line is the ITER $S_{wall} = 3.5\times 10^5$ from \req{swall}.

\begin{figure}
 \begin{center}
   \includegraphics[height=5cm]{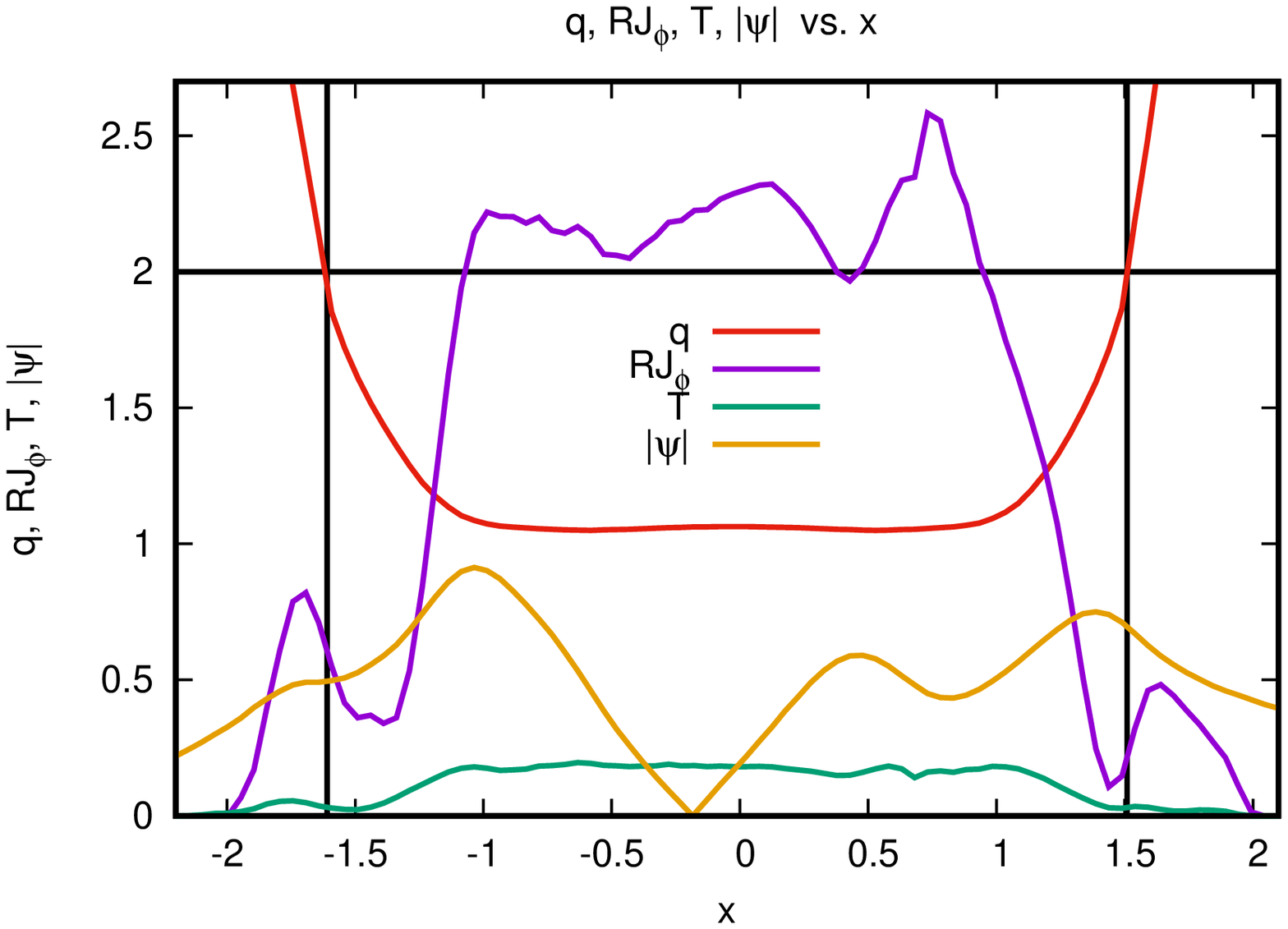}(a)
   \includegraphics[height=5cm]{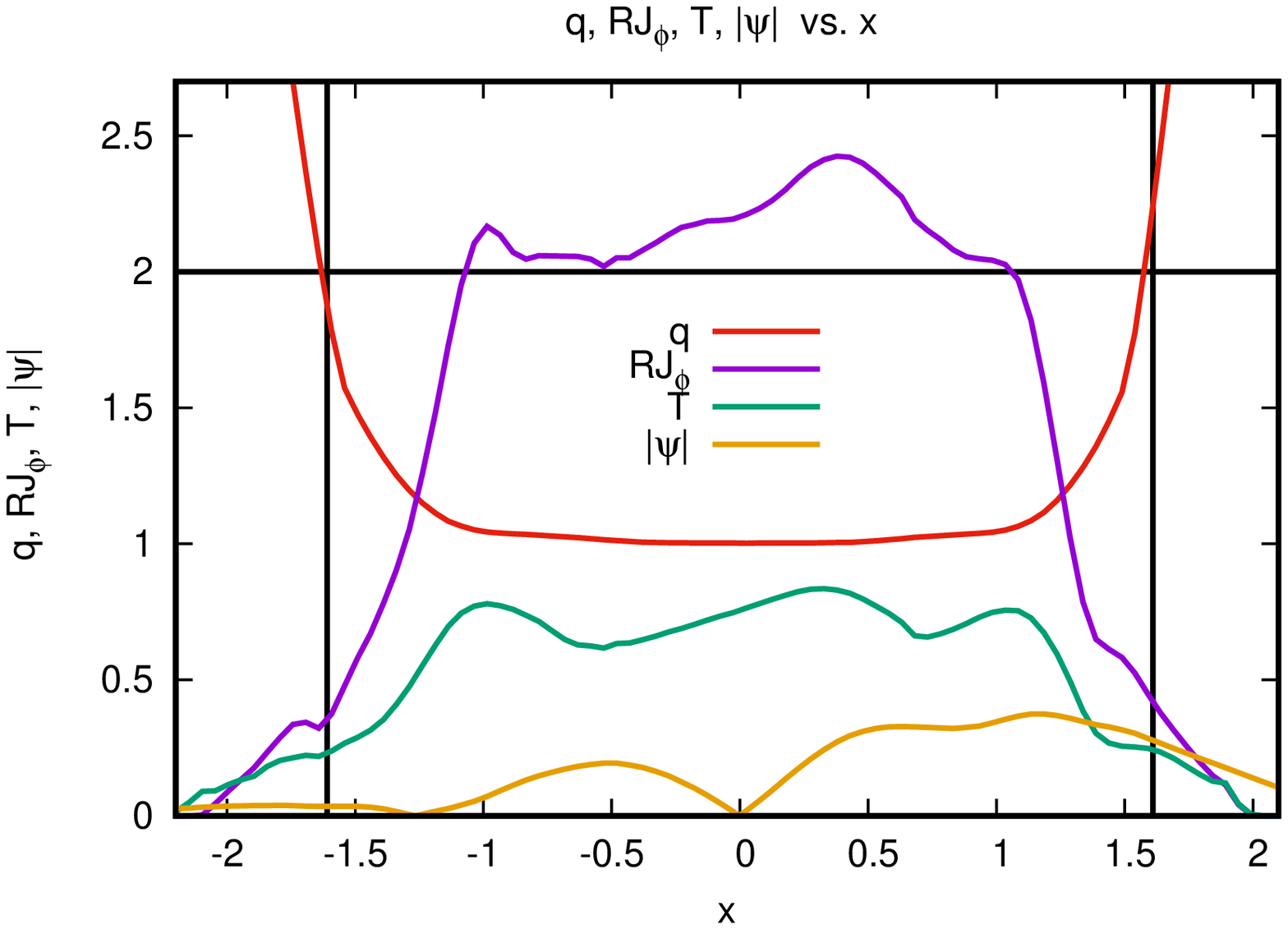}(b)
\end{center}
\caption{\it
Profiles of $q$, toroidal current density $RJ_\phi$, temperature  $T$
and toroidally
varying magnetic flux $|\tilde{\psi}|$
 as functions of $R - R_0$, with $ Z = 0.$
(a) profiles  during a RWTM, shown in \rfig{iterw3d}. 
(b) the same quantities during a disruption without a RWTM, in \rfig{iterw3d2}.
}
\label{fig:initial2}
\end{figure}

\begin{figure}
 \begin{center}
   \includegraphics[height=7cm]{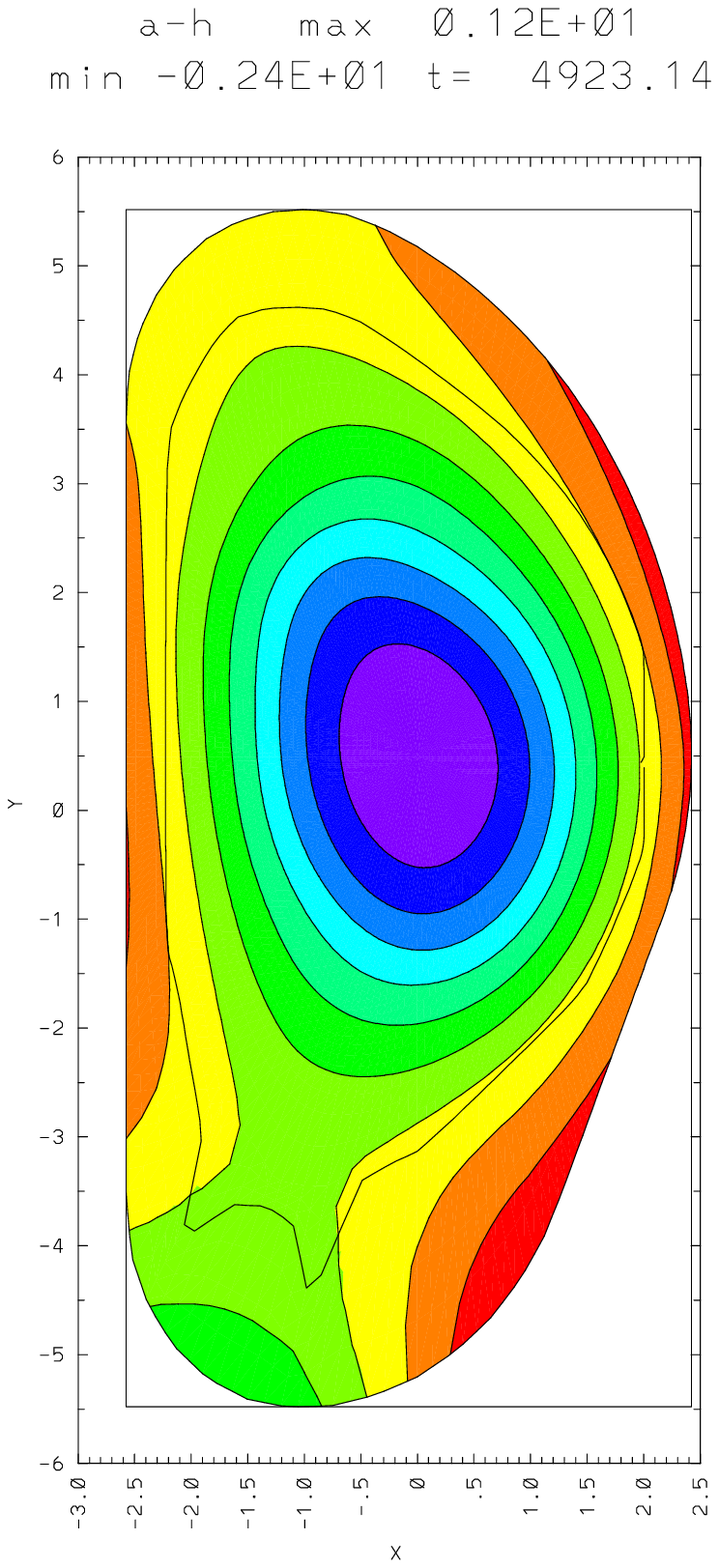}(a)
   \includegraphics[height=7cm]{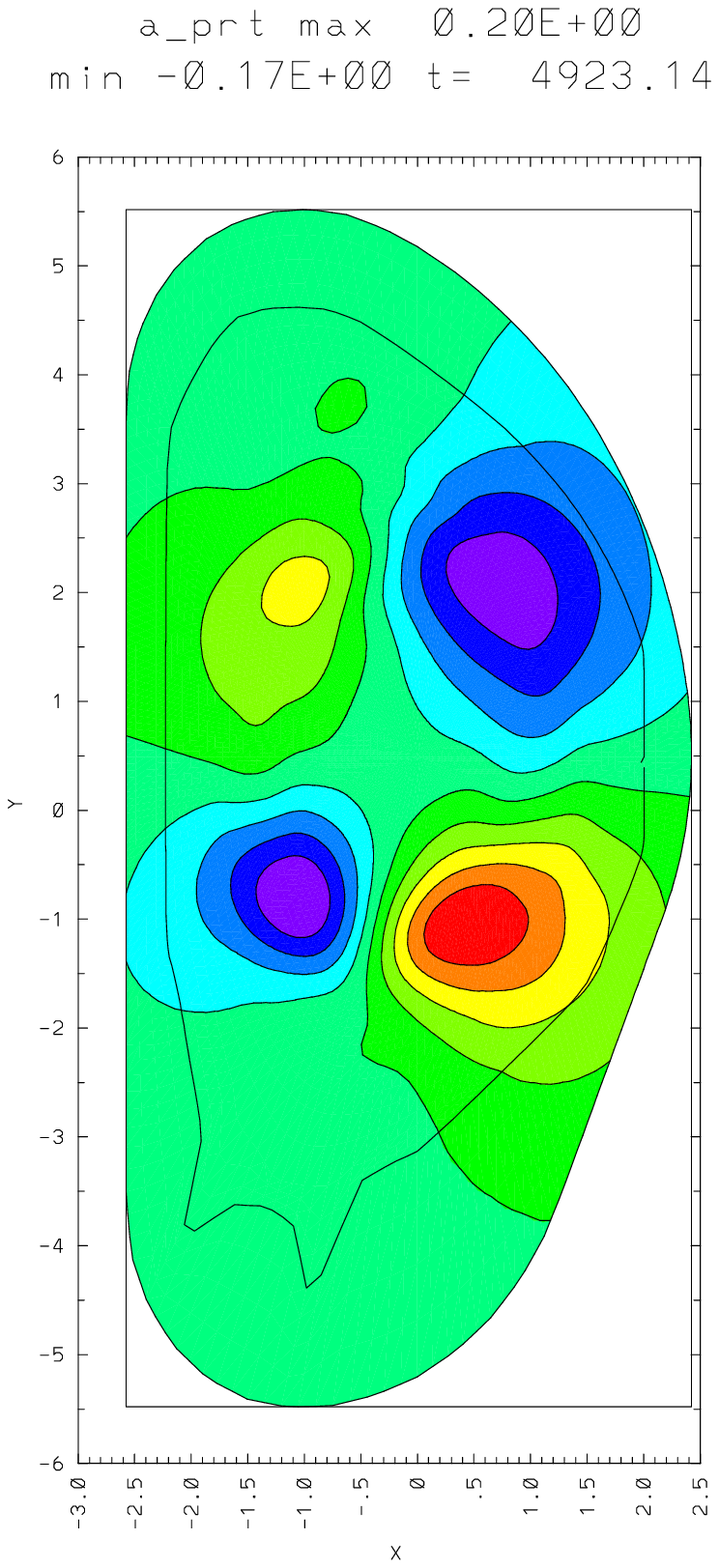}(b)
    \includegraphics[height=7cm]{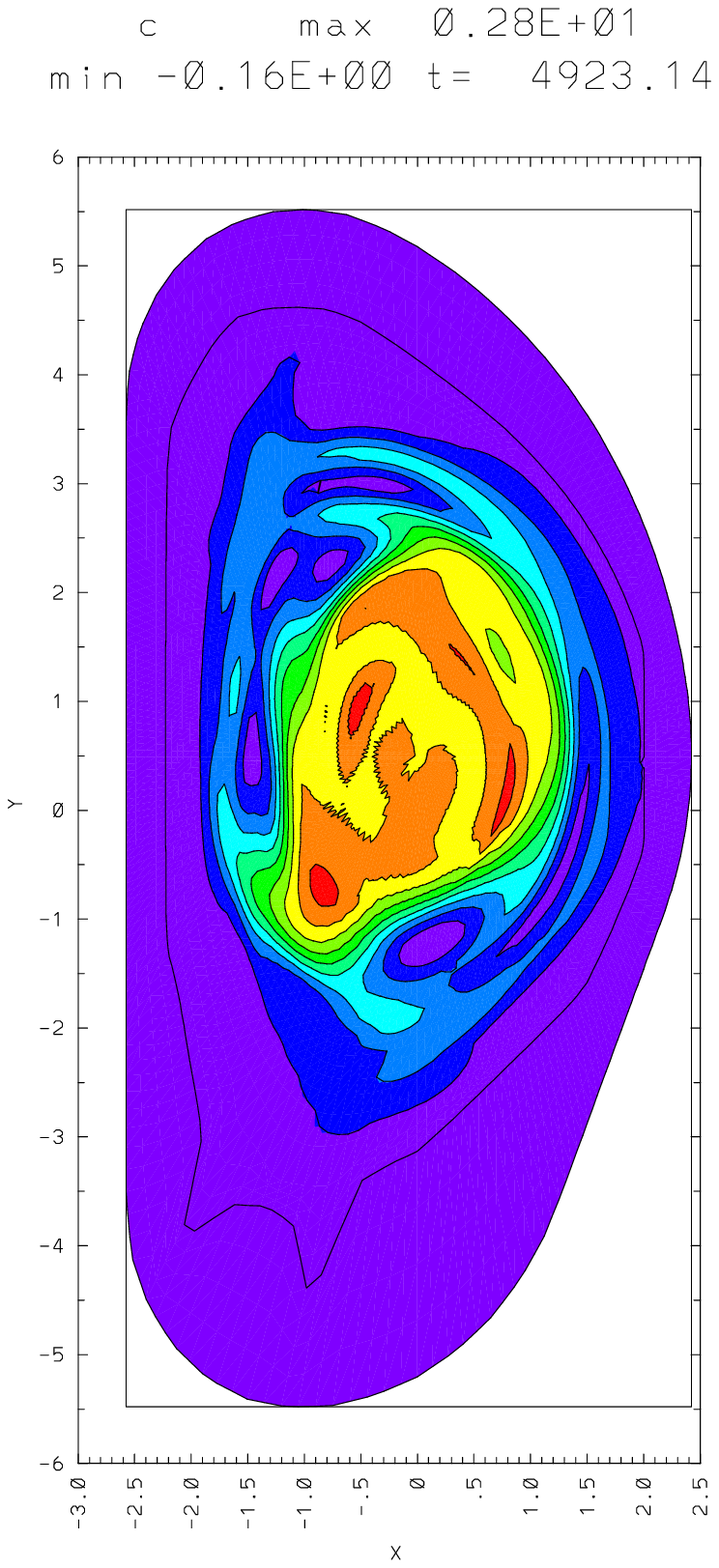}(c)
    \includegraphics[height=7cm]{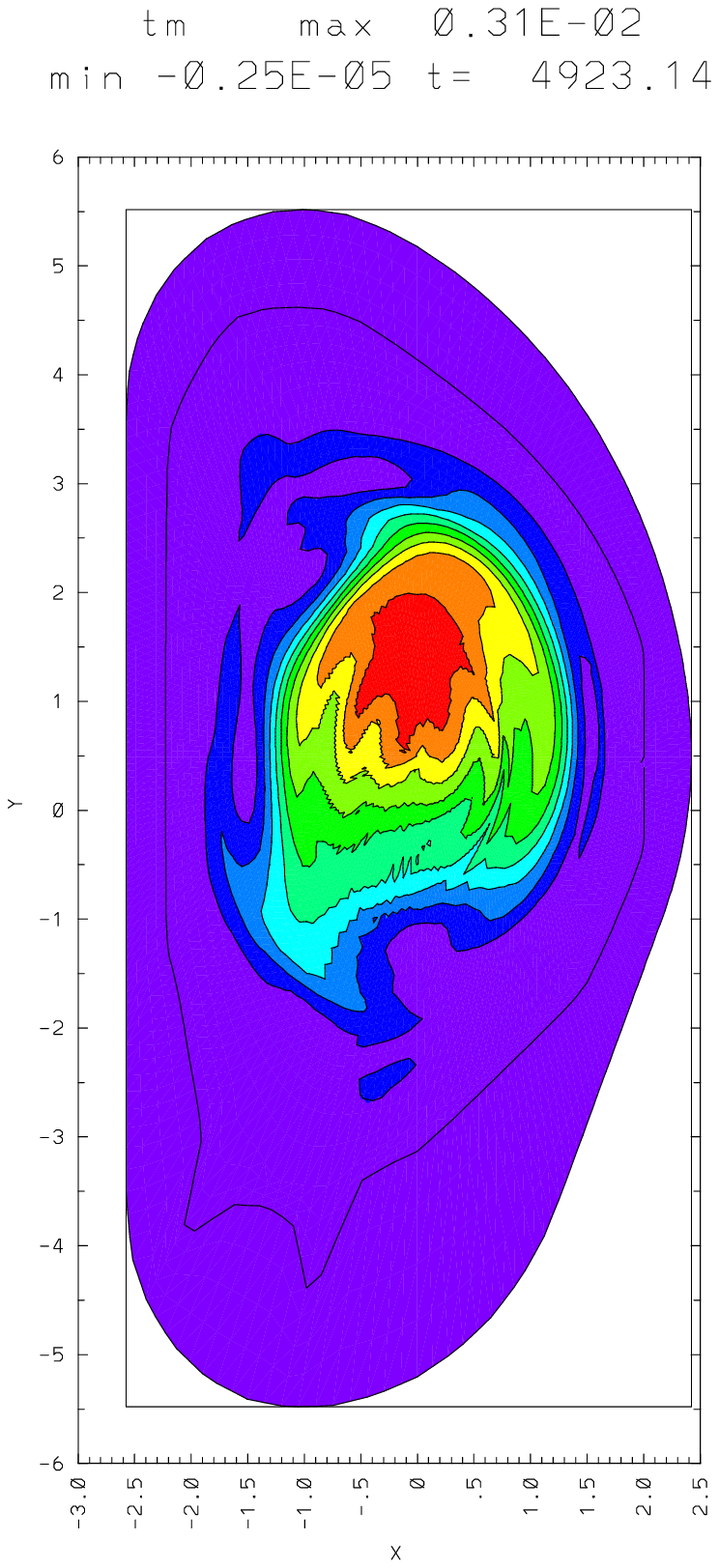}(d)
\end{center}
\caption{\it
(a) ITER simulation, $\psi$  at time $t = 4923 \tau_A,$ $S = 10^6,$ $S_{wall} = 10^3.$
(b) perturbed $\tilde{\psi}$, (c) toroidal current $J_\phi$, and
(d) $T$ at $t = 4923 \tau_A.$
The $\tilde{\psi}$ contours penetrate the outer wall.
There is a $(2,1)$ RWTM.
}
\label{fig:iterw3d}
\end{figure}

\begin{figure}
 \begin{center}
   \includegraphics[height=7cm]{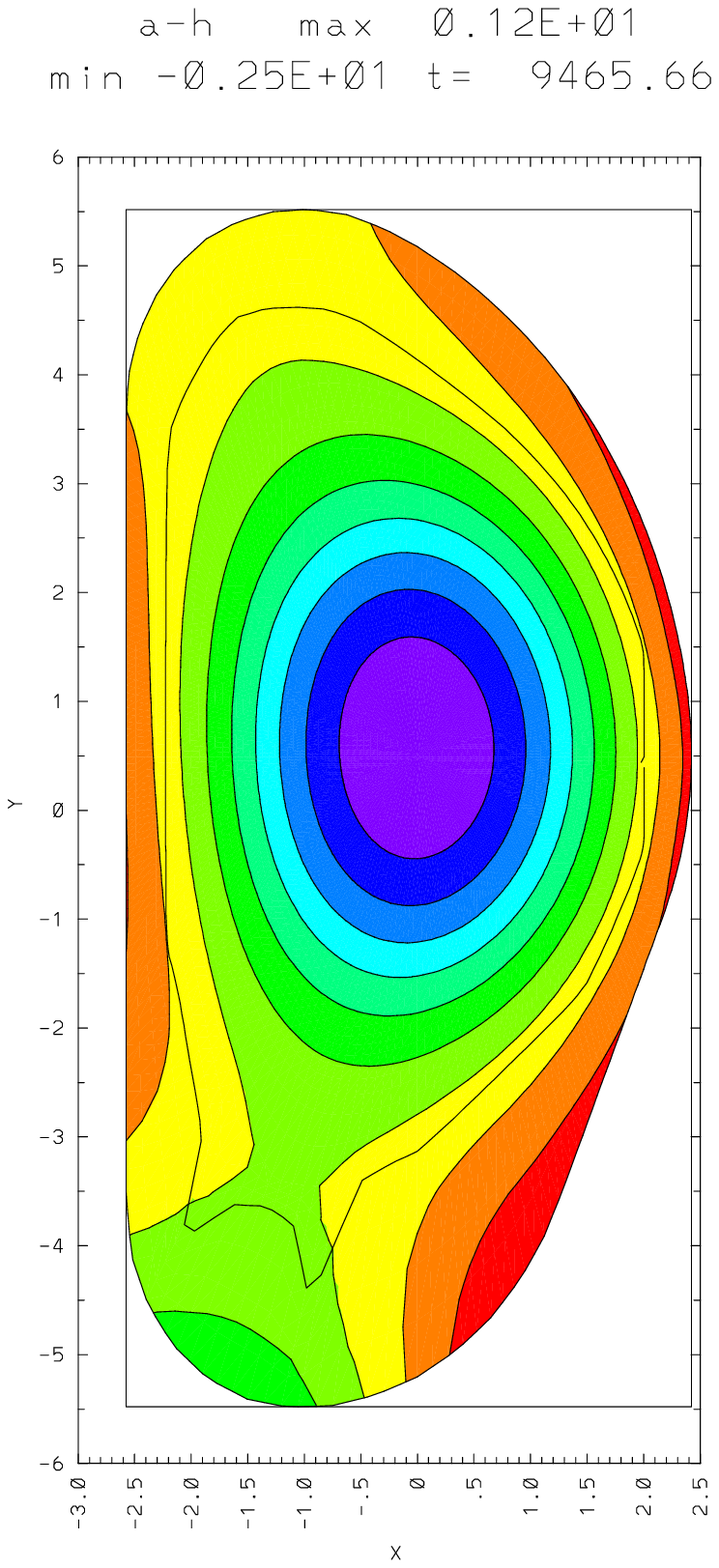}(a)
   \includegraphics[height=7cm]{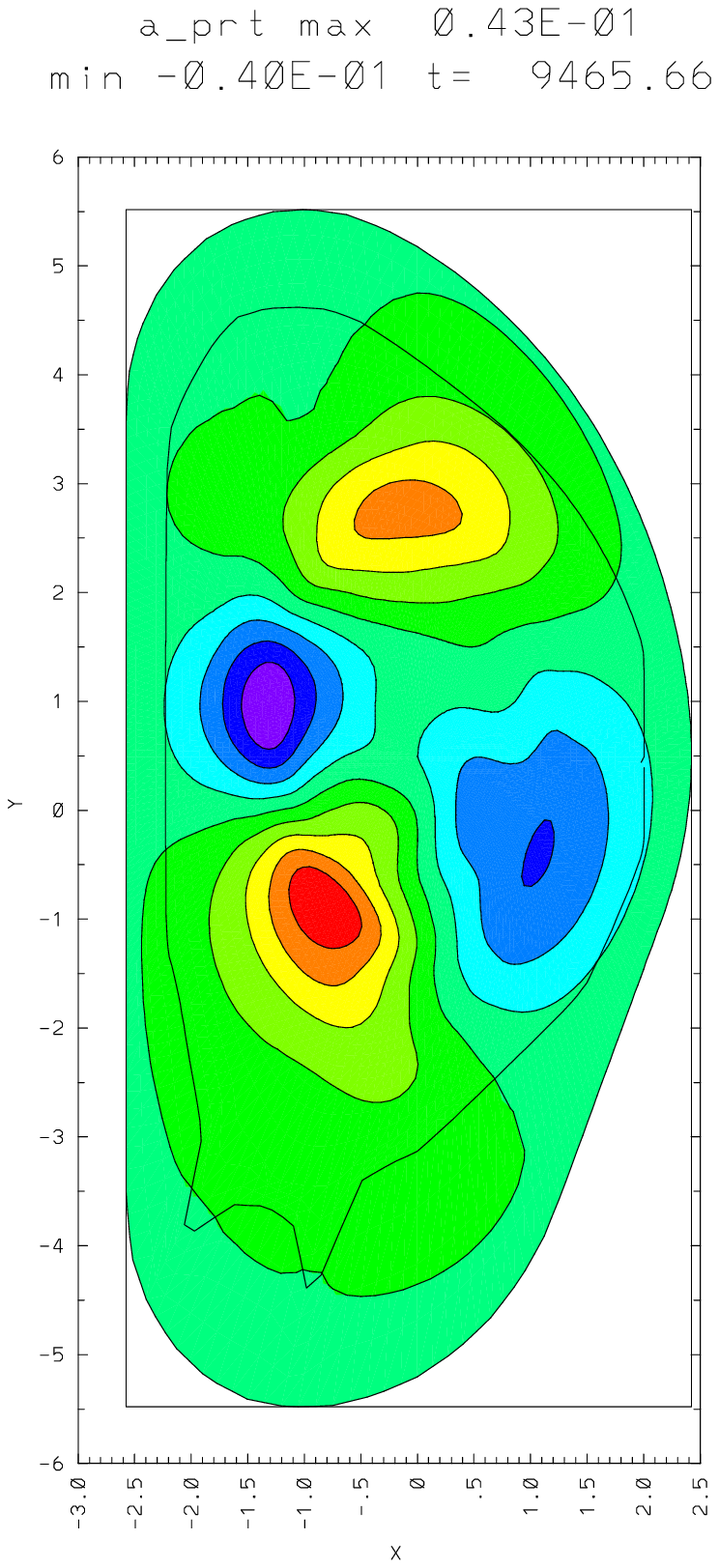}(b)
    \includegraphics[height=7cm]{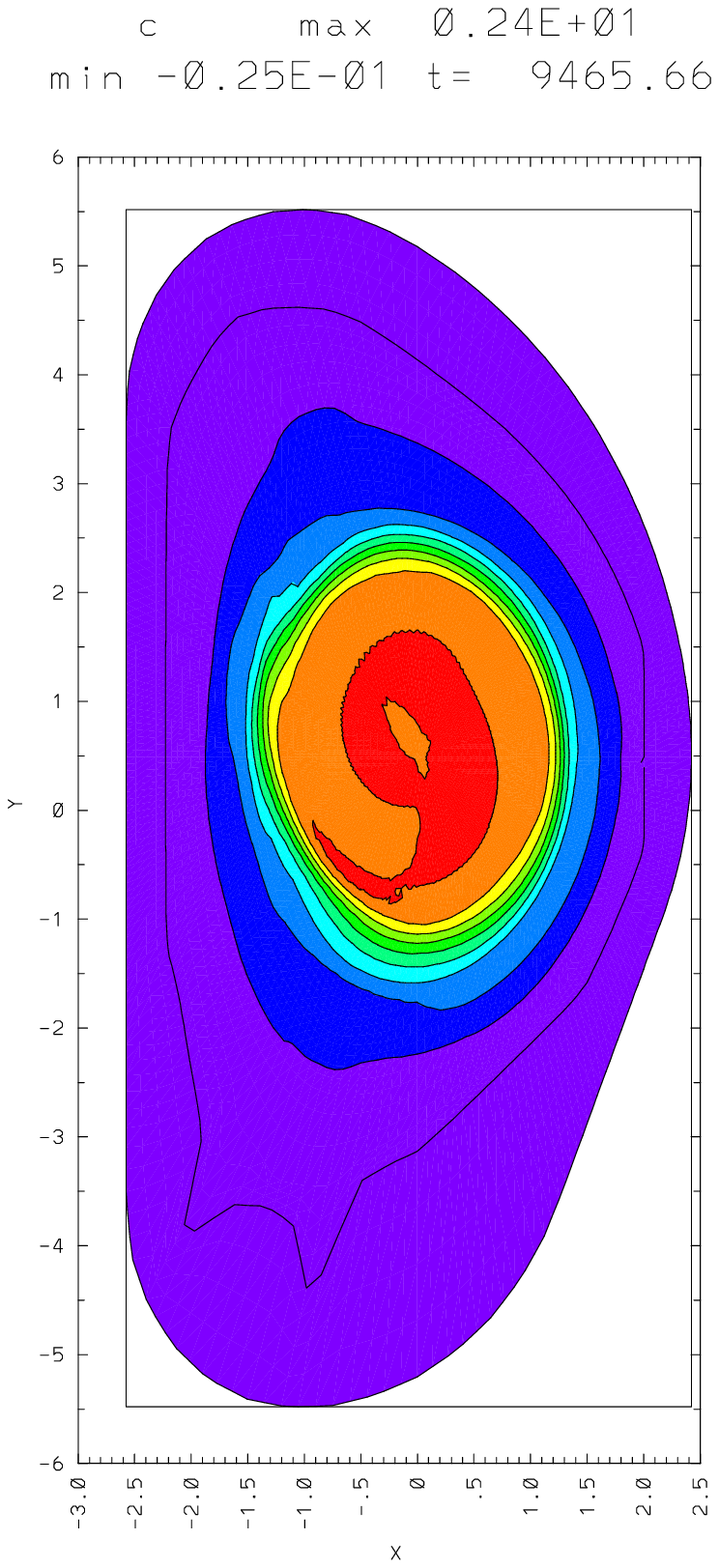}(c)
    \includegraphics[height=7cm]{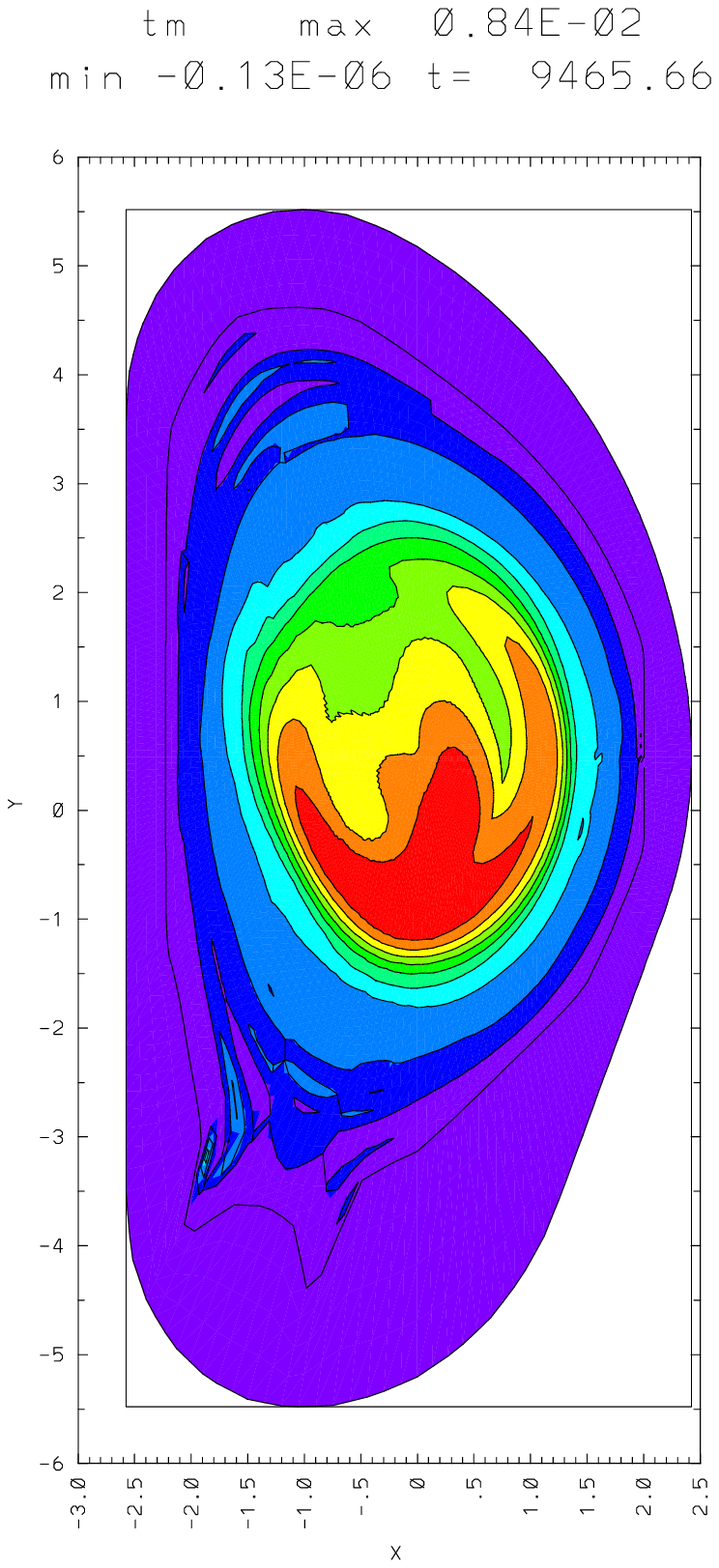}(d)
\end{center}
\caption{\it
(a) ITER simulation, $\psi$  at time $t = 9465 \tau_A,$ $S = 10^6,$ $S_{wall} = 10^4.$
(b) $\tilde{\psi}$, (c) toroidal current $J_\phi$, and
(d) $T$ at $t = 9465 \tau_A.$
The $\tilde{\psi}$ contours penetrate the outer wall only slightly.
}
\label{fig:iterw3d2}
\end{figure}

More details of the simulations are shown in \rfig{initial2}(a) and \rfig{iterw3d}.  
Profile plots are shown in \rfig{initial2}(a) for the case in \rfig{iterwall}(a) with
$S_{wall} = 10^3$, at time $t = 4923\tau_A,$ 
when the magnetic perturbations are maximum. The contours are of $q$, $RJ_\phi$, $T$
and absolute value of perturbed poloidal flux $| \tilde{\psi} |$. Comparing to \rfig{initial}, it can be
seen that $RJ_\phi$ reveals large scale flattening around the magnetic axis, with
$q = 1$ over a large radius, as well as large distortions of the current in the vicinity 
of $q = 2.$ {A large island perturbation can be seen around $q = 2.$ Here $q$ was 
calculated from the toroidally averaged magnetic field, so it does not exhibit flattening
in the island.} 
The $T$ profile has much lower values, indicating a TQ. 
The perturbed poloidal flux $| \tilde{\psi} |$ is approximately the perturbed 
radial  magnetic field,
$ b \approx ({m}/{r}) \tilde{\psi}, $ 
with $ m = 2, r = a = 2$, 
where $  \tilde{\psi} $ is normalized to the toroidal field amplitude, 
and plotted in units of $10^{-2}.$ This indicates
$b_n \approx 4 \times 10^{-3},$ consistent with \rfig{iterwall}(a).
 Contour
plots in the $(R,Z,0)$ plane are shown in \rfig{iterw3d}
for the same case. 
 Shown at time $t = 4923 \tau_A$ are (a) magnetic flux $\psi,$
(b) perturbed toroidally varying magnetic
flux   $| \tilde{\psi}|$,
 (c) current $RJ_\phi$, (d) temperature $T$. 
From \rfig{iterw3d}(b) it is clear that there is a $(2,1)$ mode in contact with
the outer wall, consistent with a RWTM. 

Simulations with $S_{wall} = 10^4$  are shown in \rfig{initial2}(b) and \rfig{iterw3d2}.
In this case the RWTM is not significant. 
Profile plots are shown in \rfig{initial2}(b) 
 at time $t = 9465\tau_A.$ 
 Comparing to \rfig{initial2}(a), the profiles of $q$ and $RJ_\phi$ are similar.
The $T$ profile has larger  values than in \rfig{initial2}(b), 
even though it is at a later time. The perturbation of the current near $q = 2$ is much
less.
The $| \tilde{\psi} |$ profile is
small at the edge, 
with $b_n \approx 1.5 \times 10^{-3}, $ consistent with \rfig{iterwall}(b).
 Contour
plots in the $(R,Z,0)$ plane are shown in \rfig{iterw3d2}
for the same case,
 at time $t = 9465 \tau_A$. 
From \rfig{iterw3d}(b) it is clear that the $(2,1)$ mode penetrates
the outer wall  only slightly.

\section{\bf Thermal Quench Theory } \label{sec:theory}

The growth rate of the RWTM is given by 
\req{gamma}.
In the previous JET simulations \cite{rwtm21}, $c_0 = 2.2.$ 
The fit to the simulations 
in \rfig{iterwall}(a) gives $c_0 = 0.51.$
In \cite{rwtm21} $c_0$ is given as
\begin{eqnarray}
 c_0 &=& 2.46  \left(\frac{ q'r_s}{q}\right)^{2/9} f^{4/9} \nonumber \\
 f &=& \frac{(r_s/r_w)^{2m} }{[1 - (r_s/r_w)^{2m}]^2} \label{eq:c0}
\end{eqnarray}
where $r_s$ is the rational surface radius $(R - R_0)$ and $r_w$ is the wall
radius.
From  \rfig{initial}, $r_s = 1.45,$  $q^\prime = 3,$
and from \rfig{iterw3d}, $r_w = 2.5.$
Using these values in \req{c0} gives $c_0 = 0.78.$
The agreement with the simulations is not unreasonable considering that
\req{c0} was derived assuming
circular cross section straight cylindrical geometry.

The simulation results can be analyzed using \cite{rwtm21}.
The two limiting dependencies of $\tau_{TQ}$ seen in
\rfig{iterwall}(b) can be obtained from a model of parallel thermal conduction.
During the TQ, heat travels along the magnetic field as
\be \frac{\pd T}{\pd t} = \frac{1}{r}\frac{\pd}{\pd r} r (\kapl  b_r^2 + \kapp)  \frac{\pd T}
{\pd r} \label{eq:dTdt}\ee
where $b_r$ is the normalized asymmetric radial magnetic field,
assuming circular flux surfaces for simplicity.
The field is assumed stochastic, so there is an average radial magnetic field.
Integrating, the total temperature is given by
\be \frac{\pd <T>}{\pd t} = a (\kapl  b_n^2 + \kapp)   T'
 \label{eq:dTdtw}\ee
where 
$ <T> = \int T r dr,$   $T' = \pd T/ \pd r$ at $r =a $,
and $b_n = b_r$ at the wall.
Assume that  $ T' / < T> = -  a^{-3}. $
The normal magnetic field at the wall is 
\be b_n =  b_{n 0} \exp({\gamma}{t} )\label{eq:dbdt2} \ee
where $b_{n 0}$ is the initial amplitude, and ${\gamma}$ is the RWTM  growth rate.

Neglecting ${\kapp}$,
 substituting for $b_n$ in \req{dTdtw} and integrating in time, from $t = 0$ to $\tau_{TQ},$
\be
   1
  =
 \frac{ \kapl b_{n 0}^2}{2 {\gamma a^2}} [ \exp(2\gamma \tau_{TQ}) - 1]
   \label{eq:dTdt2} \ee
This gives
\be {\tau}_{TQ}  = \frac{1}{2 {\gamma}} \ln \left( 1 + 2\gamma\taupzero
\right) \label{eq:log} \ee
 which has two limits,
\be
\tau_{TQ}  =              \begin{cases}
 (2 \gamma)^{-1} \ln ( 2\gamma\taupzero ) &  \gamma \taupzero >>  1  \\
  \taupzero  &  \gamma \taupzero <<  1 .
\end{cases} \label{eq:cases} \ee
where $\taupzero$ is given by \req{taupar}. 
An {\it ad hoc} fit to the simulations is given by \req{adhoc}.

\Grn{The amplitude of $b_n$ depends on $S_{wall}.$
For larger $S_{wall},$ the TQ finishes before the RWTM
has time to reach a larger amplitude.}
Let $ \gamma {\tau}_{TQ} = 1$  as in \req{adhoc}.
From \req{dbdt2}, 
\be b_n = b_{n 0}  \exp(\gamma {\tau}_{TQ}) = e b_{n 0}. \label{eq:bratio} \ee 
\Grn{This agrees with Fig. 2(a), where $ b_n \approx 3 b_{n 0} $ for
$S_{wall}  \le  10^3$, $\gamma \tau_{TQ} = 1;$
and  $b_n = b_{n 0}  \approx 10^{-3} $ for $S_{wall} \ge 10^4$,
$\gamma \tau_{TQ} \ll 1.$} 
This gives an estimate of the maximum value of $b_n$ compared to its precursor
amplitude $b_n.$
In the experimental JET example  studied in \cite{rwtm21}, the amplitude prior to
the rapid TQ termination  was
$0.4$ of the maximum amplitude, or
$b_n = 0.45 \times 10^{-3}.$
According to \req{bratio}, the precursor value of $b_n$ \Grn{in JET} was $e^{-1} = 0.37$ of the maximum.

As a check on the simulations, $\taup / \tau_A = 0.1 (a/R)^2 b_n^{-2} = 
1.1 \times 10^4$
with $b_n = b_{n 0} = 10^{-3},$ in agreement with \rfig{iterwall}(b), which is the value of $\tau_{TQ}$
when the RWTM can be neglected.

\section{\bf  Thermal Quench  Parameters } \label{sec:params}

The formula 
\req{adhoc} may be applied to examine the effect of
using realistic parameters, in particular the dependence of $\tau_{TQ}$ on $T$
and $b_n$.
Let  $T_0 = 100 eV$, $n_0 = 10^{14} cm^{-3}, $ $R_0 = 600cm.$ 
If the parallel transport is collisionless  \cite{rer} 
then $ \kapl = \pi  R  v_{e},$
where
{$v_e = 4.19\times 10^7 T(eV)^{1/2} cm/s$ is electron thermal speed.} 
If the plasma is collisional \cite{nrl}, then 
$\kapl = (2/3)\kappa_\parallel/n = 2.1 v_e^2 \tau_e ,$
{where $\tau_e = 3.44 \times 10^5 T(eV)^{3/2} n^{-1} \lambda^{-1} s$ is 
the electron collision time.} 
A combined form with both  collisionless and collisional limits  is
\be \kapl = \frac{\pi R v_e}{1 + \pi R / (2.1 v_e \tau_e)} =
\frac{7.9\times 10^{11} (T / T_0)^{1/2}}
{1 + 10.6 (T_0/T)^2[n \lambda/(n_0 \lambda_0)]} cm^2/s. \label{eq:kapl}  \ee
The condition that the approximate mean free path exceeds the connection length, 
$2.1 v_e \tau_e > \pi R$
is 
\be T > 325 \left(\frac{n \lambda}{n_0 \lambda_0}\right)^{1/2} eV \label{eq:collisionless} \ee
where the Coulomb logarithm is  $\lambda_0 = 17.$
Then from \req{taupar}
\be \taup = .051 \left(\frac{T_0}{T}\right)^{1/2} \left(\frac{b_0}{b_n}\right)^2
\left[ 1 + 10.6 \left(\frac{T_0}{T}\right)^2 \left(\frac{n \lambda}{n_0 \lambda_0}\right)
 \right] \label{eq:taup2} s \ee
where $b_0 = 10^{-3}.$
The \Alf time is
\be \tau_A = \frac{R}{v_A} = 0.73 \times 10^{-6}  
\left(\frac{n \mu }{n_0 \mu_0}\right)^{1/2} 
\frac{B_0}{B} 
s \label{eq:talf} \ee
where the ion mass ratio $\mu_0 =  m_i / m_p  = 2$, and the magnetic field is $B_0 = 5.3 T.$
The resistive diffusion time is
$\tau_R = {a^2}/{\eta} = {a^2 \omega_{pe}^2 \tau_e}/{c^2}$
giving
\be S= \frac{\tau_R}{\tau_A} =
   3.8 \times 10^6 \left(\frac{T}{T_0}\right)^{3/2} 
\left(\frac{n_0 \mu_0}{n \mu}\right)^{1/2}  
 \left(\frac{\lambda_0}{\lambda}\right)
\frac{B}{B_0}\label{eq:sval}  \ee
The wall time in ITER \cite{gribov}  is
$ \tau_{wall} = 250ms $
and 
\be S_{wall} = \frac{\tau_{wall}}{\tau_A} = 3.5 \times 10^5 \label{eq:swall} \ee
The value of ${\gamma}$ is, from \req{gamma},\req{talf},\req{sval}, 
\be {\gamma} =  146 c_0 \left(\frac{S_{w0}}{S_{wall}}\right)^{4/9}  \left(\frac{T_0}{T}\right)^{1/2} 
\left(\frac{n_0 \mu_0}{n \mu}\right)^{2/3} s^{-1}  \label{eq:gammaval} \ee 
where $S_{w0} = 3.5\times 10^5$.

In \cite{rwtm21} the growth rate of the RWTM was compared to the ideal plasma
resistive wall mode (RWM) \cite{bondeson,fitzpatrick,liu}, 
with growth rate
$\gamma_{RWM} = c_1 S_{wall}^{-1} \tau_A.$ \Grn{Taking 
$c_0=.5, c_1 = 1,$ and \req{swall} for  $S_{wall},$ the condition that
$\gamma > \gamma_{RWM}$ is $T < 1.5 KeV.$} 

\begin{figure}
 \begin{center}
   \includegraphics[height=5cm]{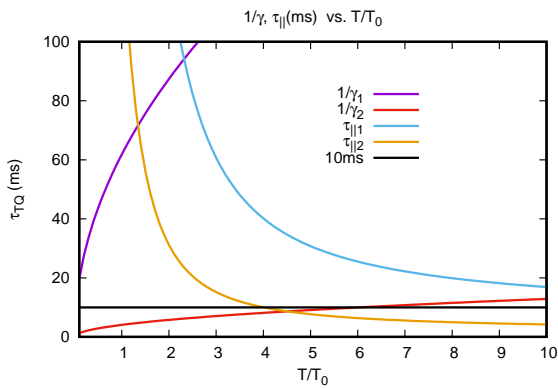}
\end{center}
\caption{\it 
$\taup$ and $1/\gamma$, where  $1/\gamma_1$ is for ITER  with $S_{wall} = 3.5\times 10^5$, 
$c_0 = .51,$ 
and  $1/\gamma_2$ is for JET, with
$ S_{wall} = 7 \times 10^3$, $c_0 = 2.2$. The
 $\taup$ values are $\tau_{\parallel 1}$ with $b_n =  10^{-3}$, 
           and $\tau_{\parallel 2}$ with $b_n =  2 \times 10^{-3}.$
The value $\tau_{TQ} = 10ms$ is also shown.
}
\label{fig:gamtau}
\end{figure}

Using \req{taup2}, \req{gammaval}, 
\rfig{gamtau}  shows $1 / \gamma$ from \req{gammaval} for $ 0.001 \le T/T_0 \le 10$. 
The curve $1 / \gamma_1$ has ITER values $S_{wall} = 3.5\times 10^5$,
$c_0 = .51,$ and
$1 / \gamma_2$ has the JET value \cite{rwtm21}  $S_{wall} =  7\times 10^3$, $c_0 = 2.2.$
Both ITER and JET have
approximately the same value of $\tau_A.$
The $\tau_\parallel$  curves are  $\tau_{\parallel 1}$ with $b_n = 10^{-3}$,
                      and   $\tau_{\parallel 2}$ with $b_n = 2 \times 10^{-3}$.
The TQ time for a given $T$ is the lesser of $1 / \gamma$
or $\taup.$
The value $\tau_{TQ} = 10ms$ is also shown.

\rfig{gamtau}  shows the important difference between ITER and JET. In ITER $\tau_{TQ}$ is determined
by $\taup,$ while in JET,  $\tau_{TQ}$ is determined by $1/\gamma,$ and is
much  less than in ITER.

It is clear from \rfig{gamtau}  that there are two different temperature regimes. In the 
collisionless regime \req{collisionless} the condition
$\taup \ge  10ms$ requires that
$T/T_0 \le  26 (b_0/b_n)^4.$ In this regime 
$\tau_{TQ}$ is very sensitive 
to $b_n.$
In the collisional regime, $ T  < 325 eV,$ the criterion is approximately 
$T / T_0 \le  4.9 (b_0 / b_n)^{4/5},$
a much weaker scaling with $b_n$.

The simulations presented here give $b_n = b_0 = 10^{-3}.$ An empirical 
 scaling of locked mode perturbation amplitudes $B_{ML}$ before the TQ  \cite{devries} 
when applied to ITER, found that for 15 MA operation with $q_{95} = 3.2$ and
internal inductance $0.9,$ the maximum value of $ B_{ML}(r_c) / B_\theta(a) $ $\approx 5 \times 10^{-3},$  where $r_c$ is the vacuum
vessel radius, with $r_c = 1.3 a.$ To measure the field at $a$, this value must be
multiplied \cite{devries}  by 
$(a/r_c)^3 = 2.2.$ 
Normalizing to the toroidal field, 
$B_\theta = B_T a/ (qR),$ this is
$  b_n \approx 5 \times 10^{-3} \times 2.2   a /( q R) = 1.2 \times 10^{-3}. $
This is approximately the same value found in the simulations.
Another estimate \cite{devries} assumes a maximum island width $w/a = 0.3.$
This gives $b_n = (r_s / a)^3 (a q^{\prime} / q)(w/a)^2 a /(16 R) \approx 2.1 \times 10^{-3},$ 
using \rfig{initial} to take the $q = 2$ rational surface at $r_s = 1.45$ 
and $a q^{\prime} /q = 3.$
These values motivate the choices of $b_n$ in \rfig{gamtau}.

\section{\bf Discussion and Conclusion } \label{sec:conclusion}

The simulations and theory  represent what might be expected in 
ITER locked mode disruptions. In present tokamaks, there is a long 
precursor phase with moderate thermal loss
caused by tearing modes. This is  followed by a rapid TQ termination, which
is seen in JET \cite{devries} , DIII-D \cite{sweeney}, and other experiments. 
It appears that the fast
termination is caused by a RWTM \cite{rwtm21}, with the timescale of the 
mode growth time. 

\Prp{It is important to 
demonstrate that locked modes
will be quite different in ITER than in present experiments.}

The RWTM growth rate scales as $\gamma \propto S_{wall}^{-4/9}.$ 
In ITER, $S_{wall}$ is $50$ times larger than in JET,
so that $\gamma$ is at least $7$ times smaller in ITER. 
The simulations imply that $\gamma$ is even smaller. 
The difference between the TQ time in ITER and JET is illustrated in
\rfig{gamtau}. In JET and perhaps in other tokamaks, the TQ time can be set by the RWTM, while
in the  ITER edge collisionless regime, it depends on parallel magnetic transport.
In the ITER edge collisional regime, the TQ time can be so long that it is effectively absent.
The plasma remains in the precursor phase.

\Grn{The $S$ values in the simulations  are
low compared to experiments. The effect is to shorten the growth times of tearing
modes and RWTMs. A  main point of the paper is to verify eqs. \req{gamma},\req{adhoc} 
 for the
TQ time. With these formulas, it is possible to calculate $\tau_{TQ}$ with realistic
parameter values, as in Fig. 6. Two parameters are needed, $c_0$ in the RWTM growth rate,
and $b_n.$ The constant $c_0$ was estimated from simulations and is also given by the
theory. The value of $b_n$ was taken from the simulations, and from the estimates in 
\cite{devries}. The value of $b_n$ comes from saturated tearing modes
in the precursor phase of the disruption. The saturated
amplitude of tearing modes depends on $\Delta'$, not on the value of $S.$}
\Grn{In particular, the saturated amplitude of the $(2,1)$ with an ideal wall
does not depend on $S$.} 
\Grn{
In present experiments, and presumably in ITER, there can be
a long precursor phase until islands overlap. The overlap criterion
is independent of $S$. 
} \Prp{
It is not
required to use realistic values in the simulation if the parameter scalings
can be 
identified. In \cite{rwtm21,fec20,jet2017}, simulations used $S = 10^6$ and the results
were in good agreement with JET experimental data.
}

\Prp{
The disruptions discussed here are similar to
locked modes in JET and other experiments with $q_0 \stackrel{>}{\sim}  1$.
At higher $q_0,$ the $q = 2$ rational
surface moves closer to the magnetic axis.
The RWTM growth rate depends on the  quantity $f$ in \req{c0}. 
As $r_s/r_w$ decreases,  $f$
decreases as $(r_s/r_w)^{2m}.$
Only for a $(2,1)$ rational surface near the plasma edge, can there be a large wall
interaction. 
For a mode with  $m = 3$, the wall interaction is small.
Without a significant wall  interaction, the growth rate is small. 
}

\Prp{Neoclassical tearing modes (NTMs) can contribute to internal disruptions
\cite{iter}, but are not known to cause major disruptions. They are expected to
have smaller island widths in ITER than in present experiments \cite{iter}.}

\Prp{
The present paper is concerned with an ITER inductive scenario.
ITER advanced scenarios are planned with high $\beta_N,$  reversed shear,
central $q_0 > 1,$ and low $li.$ 
In high $\beta_N$  ITER scenarios,
the plasma  is  unstable to a kink mode at the ideal wall limit.
The assumption is that
plasma heating causes the no - wall limit to be reached at a lower $\beta_N$ 
than the ideal wall limit.
Between the no - wall and ideal wall limits, the plasma is unstable to a RWM,
with a long growth time.
}

\Prp{There are other possible mechanisms for a TQ, 
which are not included in the present study. 
In particular there are
asymmetic vertical displacement events
(AVDEs)  \cite{fec20, jet2017},
 which are typically triggered by a TQ, with timescale of order of the
resistive wall time.
There are also
effects of radiation \cite{izzo,ferraro,nardon}, and density limit disruptions, which may also be an effect
of radiation \cite{greenwald}.}

The ITER edge collisional regime gives a window of very long $\tau_{TQ}.$
Accessing this regime 
would mitigate the requirements for the ITER disruption mitigation system
and runaway electron avoidance.
 \rfig{gamtau} shows that if $T = 300eV$ and $b_n \approx 10^{-3}$ 
at the edge, $\tau_{TQ}$ is
limited by the RWTM to almost $100ms.$ 

At higher edge temperatures,  or higher $b_n$,
if mitigation is required,
 it might be possible to cool the edge radiatively, to
access the collisional edge  regime, without the need for cooling the plasma interior.

\noindent{\bf Acknowledgment} {
Work supported by USDOE grant DE-SC0020127.}

\noindent{\bf Data availability statement}
The data that support the findings of this study are available from the corresponding author upon reasonable request.


\begin{thebibliography}{11}
\bibitem{devries}
P.C. de Vries, G. Pautasso, E. Nardon, P. Cahyna, S. Gerasimov,
J. Havlicek, T.C. Hender, G.T.A. Huijsmans, M. Lehnen, M. Maraschek, T. Markovic,
 J.A. Snipes and the COMPASS Team, the ASDEX Upgrade Team and JET Contributors,
Scaling of the MHD perturbation amplitude required to trigger a disruption 
and predictions for ITER,
Nucl. Fusion \textbf{56}  026007 (2016)
\bibitem{sweeney} 
R. Sweeney, W. Choi, M. Austin, M. Brookman, V. Izzo, M. Knolker, R.J. La Haye, A. Leonard,
 E. Strait, F.A. Volpe and The DIII-D Team,
Relationship between locked modes and thermal quenches in DIII-D,
Nucl. Fusion 58 (2018) 056022 

\bibitem{rwtm21}
H. Strauss and JET Contributors, 
Effect of Resistive Wall on Thermal Quench in JET Disruptions,
Phys. Plasmas \textbf{28}, 032501 (2021); doi: 10.1063/5.0038592.

\bibitem{iter}
T. Hender, J. C. Wesley, J.  Bialek, A.  Bondeson, A. Boozer, R. J.
Buttery, A.  Garofalo, T. P Goodman, R. S.  Granetz, Y.  Gribov, O.  Gruber, M.
 Gryaznevich, G.  Giruzzi, S.  G\"unter,
  N.  Hayashi, P.  Helander, C. C.  Hegna, D. F.  Howell,
 D. A.  Humphreys, G. T. A.  Huysmans, A.
W.  Hyatt, A.  Isayama, S. C.  Jardin, Y.  Kawano, A.  Kellman, C.
 Kessel, H. R.  Koslowski, R. J.  La Haye, E.  Lazzaro, Y. Q.  Liu, V.
  Lukash, J.  Manickam, S.  Medvedev, V.  Mertens, S. V.  Mirnov, Y.
Nakamura, G.  Navratil, M.  Okabayashi, T.  Ozeki, R.
 Paccagnella, G.  Pautasso, F.  Porcelli, V. D.  Pustovitov, V.  Riccardo, M.  Sato, O.
 Sauter, M. J.  Schaffer, M.  Shimada, P.  Sonato, E. J.  Strait, M.
Sugihara, M.  Takechi, A. D.  Turnbull, E.  Westerhof, D. G.  Whyte, R.
Yoshino, H.  Zohm and the ITPA MHD, Disruption and Magnetic Control
Topical Group, 
Progress in the ITER Physics Basis,
MHD stability, operational limits, and disruptions (chapter 3)
Nuclear Fusion \textbf{47} S128 - 202 (2007).
\bibitem{lehnen}
M.Lehnen, K.Aleynikova, P.B.Aleynikov D.J.Campbell, P.Drewelow, N.W.Eidietis,
Yu.Gasparyan, R.S.Granetz, Y.Gribov, N.Hartmann, E.M.Hollmann,  V.A.Izzo, S.Jachmich,
S.-H.Kim, M.Kočan, H.R.Koslowski, D.Kovalenko, U.Kruezi, A.Loarte, S.Maruyama,
G.F.Matthews, P.B.Parks, G.Pautasso, R.A.Pitts, C.Reux, V.Riccardo, R.Roccella,
J.A.Snipes, A.J.Thornton, P.C.de Vries, EFDA JET contributors,
Disruptions in ITER and strategies for their control and mitigation,
Journal of Nuclear Materials, \textbf{463}, 39 (2015)

\bibitem{izzo}
 V. A. Izzo, D. G. Whyte, R. S. Granetz, P. B. Parks,
E. M. Hollmann, L. L. Lao, J. C. Wesley,
Magnetohydrodynamic simulations
of massive gas injection int Alcator C - Mod and DIII-D plasmas,
Phys. Plasmas \textbf{15}, 056109 (2008).
\bibitem{ferraro} N.M. Ferraro, B.C. Lyons, C.C. Kim, Y.Q. Liu and S.C. Jardin,
3D two-temperature magnetohydrodynamic modeling of fast thermal quenches due to injected impurities in tokamaks,
Nucl. Fusion \textbf{59} (2019) 016001.
\bibitem{nardon} E. Nardon, A. Fil, M. Hoelzl, G. Huijsmans and JET contributors,
Progress in understanding disruptions triggered by massive gas injection via 3D non-linear MHD modelling with JOREK,
Plasma Phys. Control. Fusion \textbf{59}  014006 (2017).
\bibitem{reux}
C. Reux, V. Plyusnin, B. Alper,  D. Alves, B. Bazylev, E. Belonohy,
A. Boboc, S. Brezinsek, I. Coffey, J. Decker, P. Drewelow, S. Devaux, P.C. de Vries,
A. Fil, S. Gerasimov, L. Giacomelli, S. Jachmich,
E.M. Khilkevitch, V. Kiptily, R. Koslowski, U. Kruezi, M. Lehnen,
I. Lupelli, P.J. Lomas, A. Manzanares, A. Martin De Aguilera,
G.F. Matthews, J. Mlynai, E. Nardon, E. Nilsson, C. Perez von Thun, V. Riccardo,
F. Saint-Laurent, A.E. Shevelev, G. Sips, C. Sozzi1 and JET contributors,
Runaway electron beam generation and mitigation during disruptions at JET-ILW,
Nucl. Fusion \textbf{55}  093013 (2015)

\bibitem{finn95} John A. Finn,
Stabilization of ideal plasma resistive wall modes in cylindrical
geometry: the effect of resistive layers,
Phys. Plasmas 2, 3782 (1995)
\bibitem{m3d}
 W. Park, E.  Belova, G. Y.   Fu, 
X.  Tang, H. R.  Strauss, L. E.  Sugiyama,
Plasma Simulation Studies using Multilevel Physics Models,
  Phys. Plasmas \textbf{6} 1796 (1999).

\bibitem{iter2018}
H. Strauss,
Reduction of asymmetric wall force in ITER disruptions by current quench,
Physics of Plasmas \textbf{25} 020702 (2018).

\bibitem{fec20}
H. Strauss,  E. Joffrin, V. Riccardo, J. Breslau, R. Paccagnella, G.Y. Fu,
 and JET contributors,
Reduction of asymmetric wall force in JET and
 ITER disruptions including runaway electrons,
Phys. Plasmas \textbf{27}  022508 (2020)

\bibitem{schuller}
F.C. Schuller, Disruptions in tokamaks, Plasma Phys. Controlled Fusion \textbf{37}, A135 (1995).
\bibitem{pucella} G. Pucella, P. Buratti, E. Giovannozzi, E. Alessi,
F. Auriemma, D. Brunetti, D. R. Ferreira, M. Baruzzo,
D. Frigione, L. Garzotti, E. Joffrin, E. Lerche, P. J. Lomas, S. Nowak, L. Piron,
F. Rimini, C. Sozzi, D. Van Eester, and JET Contributors,
Tearing modes in plasma termination on JET:
the role of temperature hollowing and edge cooling,
 Nucl. Fusion \textbf{61} 046020 (2021)


\bibitem{gribov} Y. Gribov and V. D.  Pustovitov,
Analytical study of RWM feedback stabilisation with application to ITER,
 Proc. 19th IAEA Fusion Energy Conf. (Lyon, 2002) CT/P-12
http://www-pub.iaea.org/MTCD/publications/PDF/csp\_019c/pdf/ctp\_12.pdf



\bibitem{rer} A. B. Rechester and M. N. Rosenbluth, Phys. Rev. Lett. 40, 38 1978

\bibitem{nrl} J. D. Huba, NRL Plasma Formulary, Naval Research Laboratory
(Washington DC) 2007

 \bibitem{bondeson} A. Bondeson and D. J. Ward,
Stabilization of external modes in tokamaks by resistive
walls and plasma rotation,
Phys. Rev. Lett. 72, 2709 (1994).
\bibitem{fitzpatrick} Richard Fitzpatrick,
A simple model of the resistive wall mode in tokamaks,
Phys. Plasmas \textbf{9} 3459 (2002).
\bibitem{liu}
Fabio Villone, Yueqiang Liu, Guglielmo Rubinacci and Salvatore Ventre,
Effects of thick blanket modules on the resistive wall modes stability in ITER,
Nucl. Fusion 50 (2010) 125011.




\bibitem{greenwald}
Martin Greenwald, J.L. Terry, S.M. Wolfe, S. Ejima, M.G. Bell, S.M. Kaye, G.H. Neilson,
A new loook at density limits in tokamaks,
Nucl. Fusion \textbf{28}, 2199 (1988).
\bibitem{jet2017}
H. Strauss, E. Joffrin, V. Riccardo, J. Breslau, R.  Paccagnella,
and JET Contributors,
Comparison of JET AVDE disruption data with M3D  simulations
 and implications for ITER,
Phys. Plasmas \textbf{24} 102512 (2017).


\end{thebibliography}
\end{document}